# Pro-f-quiz: increasing the PROductivity of Feedback through activating QUIZzes

**Kris Aerts[1], Wouter Groeneveld[1]**
[1]OVI@ACRO, Department Computer Science, KU Leuven Campus Diepenbeek, Belgium

*Abstract*

*Feedback beyond the grade is an important part of the learning process. However, because of the large student groups, many teachers in higher education are faced with practicalities such as the limited time to prepare and communicate the feedback to individual students. We have set up an experiment, titled Pro-f-quiz, in which over two years 236 students participated and in which the feedback is communicated through an online quiz that activates the students to reflect upon their solution. The system can be set up in a very limited time compared to booking individual time slots. The results show that approximately 85% of the students appreciate the approach with 60% indicating that they reflect more intensively about their work than when the feedback is transmitted traditionally. Moreover, the grade of the students participating in the project was substantially higher than students not participating.*

**Keywords:** *grading; feedback; quiz; reflection; student appreciation.*

*Pro-f-quiz: increasing the **pro**ductiveness of **f**eedback through activating **quiz**zes*

## 1. Introduction

Feedback is important in the learning process of students. Providing proper feedback–not overly negative but constructive, enthusing and with a focus on reflection–is of paramount importance but comes with challenges such as tight time constraints and a perceived limited return on investment. Time is a many-facetted problem. First, the teacher's time: to grade, to formulate and to communicate the feedback. Formulating feedback takes extra time over grading because the match with the evaluation criteria must be made explicit and the required improvement elucidated. Communicating can also take significant time, especially when done orally. The second and often underestimated issue is the student's time. Fitting the schedule in both the teachers' and the students' agenda would imply an extended time span between the first and the last student getting their feedback. In terms of equal treatment this poses a major challenge when the idea of the feedback is to be formative, i.e., helping them in performing better at consecutive tasks.

The limited return on investment is the other problem: Crisp (2007) reports a feeling of wasted time providing feedback, as also confirmed by individual teachers in our environment. Our conjecture is that students, unless triggered otherwise, act as mere receivers of feedback and often prefer to scan the feedback superficially, only searching for the actual grade.

This paper presents **Pro-f-quiz,** a **productive** methodology for giving **feedback** through means of a **quiz**. The quiz consists of concrete, multiple choice questions closely tied to the task at hand. After each question, the student is provided feedback, not only the correct answer but also about the imperfections of the incorrect options. Only after finishing the entire quiz, the students can see their grade and score on each of the questions. We believe this methodology improves the productivity of the feedback process because students are triggered to reflect about their own solution before being able to read the answer, while only requiring a limited time effort for the teacher. The experiment has been performed in the first bachelor year of Engineering Technology for in total 374 students, of which 236 students, divided in 146 project groups, voluntarily participated.

We defined the following research questions to evaluate the productivity of the methodology:

- **RQ1: Do students appreciate the methodology?**
- **RQ2: Does it increase reflection?**
- **RQ3: Is the time needed to set up the feedback system limited in comparison to traditional approaches?**

The remainder of this paper is as follows: Section 2 provides related work; we describe the context and setup of the experiment in Section 3 while presenting the results and the lessons learnt in Section 4. After Section 5, exploring the threats to validity, Section 6 concludes with a summary and suggestions to expand the experiment.

## 2. Related Work

The use of (automated) assessment and feedback systems to facilitate students' learning has been extensively researched in recent years. Barriocanal et al. (2002) found out that a straightforward application of unit testing, a fully automated feedback mechanism for testing software, does not improve student's engagement. Chatzopoulou (2010) proposes "adaptive assessment", where students with different abilities are served different sets of questions. Matthews et al. (2012) introduced such a system and show that it is especially interesting for a resource-based learning approach that focuses on process assessment. The framework allows for custom comments by instructors. The quantity of feedback and response time was improved, but the quality rates dropped slightly.

Sherman et al. (2013) found out that computer-assisted automated grading systems positively affect student feedback and response time, allowing educators to quickly grade multiple complex assignments while at the same time allowing students to increase their submission per assignment rate. More recently, Thangaraj et al. (2022) compared six recent feedback systems for introductory programming courses in higher education on five different axes. They praise the immediate feedback and the opportunities for the student to initiate the feedback by submitting a work, but also conclude that more development is needed to be more effective in motivating the learning process.

When it comes to the format of feedback, Funk and van Diggelen (2014) claim that although written feedback is still used a lot, students are dissatisfied with the way they are formulated and the quality in general. A two-legged categorical system should be developed to mitigate the fluctuating quality, they conclude. The first category, focusing on the content of learning, should cover task, proces, vision, identity, etc. The second category determines the form of giving feedback: positive-negative, specific-general, limited-elaborate, etc. A quiz-like feedback format as presented in this work avoids most written feedback shortcomings outlined by Funk and van Diggelen.

Online quiz systems have been successfully employed to support educators in providing feedback in large-class courses. Furthermore, most of them allow for multiple attempts, resulting in a significant increase in students' quiz scores. Also, Mendoza and Lapinid (2022) discovered that the positive attitude of students towards feedback increased compared to conventional text-based feedback systems, as it allowed them to better manage their time and help them understand the theoretical material at hand. Although Mendoza and Lapinid did not use the system to effectively grade students, they do note that average results of online quizzes is a good predictor of final exam grades, as also unveiled by Cohen and Sasson (2016) and McDaniel et. al. (2007), who conclude that in the classroom testing can be used to promote learning, not just to evaluate learning. Our approach aims to combine both.

## 3. Experimental Setup

The experiment took place during two consecutive years in the course *Software design in Java* (4 ECTS) at the joint program 'Engineering Technology' at UHasselt and KU Leuven Campus Diepenbeek (Belgium). The course, an *"Objects First"* course (Cooper (2003), Barnes et. al. (2006)), is attended by roughly 200 students yearly. It is situated in the initial general engineering phase of three semesters (1,5 year), after which students choose one of seven options ranging from Chemistry or Construction to Electronics-ICT or Software Systems. All students must pass the course. As students interested in majors outside IT often demonstrate less interest, we use nudging techniques as stimulation. One form used to be an early but non-committal task–no grades, only feedback–that helps them in succeeding in the consecutive assignment (worth 33% of the grade). Because the number of students who took up the challenge gradually decreased to less than 10, we changed the task from non-committal to voluntary, and started grading the tasks, using a bonus system to ensure that students are never punished for performing poorly in the voluntary task. Participation increased to 60-70% but giving individual high-quality feedback would now require a lot of effort.

To minimize the effort and maximize the effect we set up of first version of the experiment in 2021-'22. Instead of directly distributing the feedback and scores, we composed a quiz of 10 multiple-choice questions named the **Pro-f-quiz**. Some questions were basic: *"Did you declare your data members as public?"*; some more introspective: *"Wat representation did you use for the quality of the snow?"*; other questions listed options that could or could not be applicable. Each question was followed by immediate feedback explaining the rationale, e.g. *"A String is not suited because **any** text can be used, not only the values we enumerated."* We also provided an anonymized Excel-sheet with hashed student names containing personalized feedback and grades. Student can only retrieve their hash by finishing the quiz. We also added an additional question to the quiz *"Did you learn from this feedback quiz?"* Because 87% answered positively, and because we later noticed that the grade of the students who did the voluntary task was 3,63/20 higher than other students, we conducted a second version of the experiment in 2022-'23 with additional questions regarding the feedback mechanism and how much effort they put in the voluntary task. Results were collected through the learning platform Toledo, a derivate of Blackboard and analyzed in Excel.

## 4. Results and Discussion

*Table 1* shows a high overall participation rate of 63% of the students doing the voluntary assignment. The participation rate, counted in number of quizzes completed, increased from 36% to 48,4% in the second year. This implies an average error rate of 7,53% with a confidence level of 95%. If we consider the response as representative for the entire project group, the error rate drops below 6%, thus considering the sample as highly reliable.

*Kris Aerts, Wouter Groeneveld (authors)*

**Table 1 Participation in the experiment**

| Year | Students | Projects | Students with project | Completed quizzes |
|---|---|---|---|---|
| 2021-'22 | 192 | 69 | 127 | 69 |
| 2022-'23 | 182 | 77 | 109 | 88 |

On the overall question "*Did you learn from this feedback quiz?*" 87,0% of the students responded positively in the first year and 83,7% in the second, which is consistently good.

Given an expected working time of approx. 4 hours and the responses on question 12 "*How much time did you spend in the task*", we divided the population as follows:

- Students working shorter than expected on the task (2 to 3 hours): 14 students
- The group that worked as expected (3,5 to 5 hours): 26 students
- The group that took slightly more time (5,5 to 8 hours): 24 students
- Students who worked a lot more (8,5 up to 20 hours): 20 students

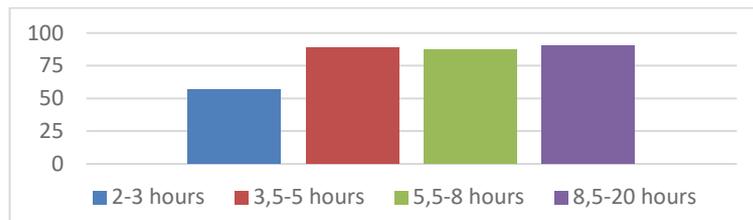

*Figure 1 Percentage of students who learnt from the quiz, versus time spent on the task*

*Figure 1* shows a distinct difference between the groups that had to work the least versus the most. The first group attributes to 16,7% of the population but contains almost half of the students who did not learn from the feedback: their approval ratio is 57,1% as opposed to 90% in the last group, which is the more important target group. As they had to work much harder to achieve a similar result, they can probably benefit more from high quality feedback. *Figure 2* shows the results classified according to their grade on the task. The percentage of positive responses reaches 100% in the group who performed the least. Together with *Figure 1*, this results in a positive answer on **RQ1** *(Do students appreciate the methodology?)*.

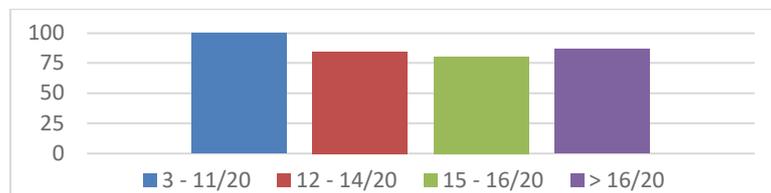

*Figure 2 Percentage of students who learnt from the quiz, versus grade on the task*

*Pro-f-quiz: increasing the **pro**ductiveness of **f**eedback through activating **qui**zzes*

The mere observation that students benefit from the feedback is insufficient to claim that this methodology is superior. Therefore, we explicitly asked the following question: "*To what extent has this way of feedback triggered you to reflect more on the quality of your task?*" ***Table 2*** shows that 60% of the students is convinced that the approach with the quiz helped them in reflecting more about the quality of their task.

**Table 2 Distribution of students reflecting more than when directly given the feedback**

| Did you reflect more than when directly given feedback? | Students | Percentage |
|---|---|---|
| Very definitely yes | 28 | 32% |
| Probably yes | 24 | 28% |
| Undecided | 28 | 32% |
| Probably not | 4 | 5% |
| Very definitely not | 3 | 3% |

The final two questions look at the attitude of the students. *Table 3* shows positive intentions of the students that are not aligned with past observations: 61 students claim that they would make the task irrespective of the promise of a reward, whereas this shrunk to no more than 10 students when there was no reward. *Table 4* confirms to a certain degree the presumption put forward in the introduction that many students are only interested in the grade, but on the other hand more than 67% of the students in the sample value the feedback at least equally important as the grade. The effort put in providing is thus highly appreciated by the students.

**Table 3 Doing the task without reward**

| Would you have done the voluntary task without bonus reward? | Students |
|---|---|
| Very definitely yes | 23 |
| Probably yes | 38 |
| Undecided | 13 |
| Probably not | 10 |
| Very definitely not | 3 |

**Table 4 Appreciation of feedback**

| Feedback versus grade | Students |
|---|---|
| I only remember the grade. | 8 |
| Grade is more important than feedback. | 20 |
| Grade & feedback are equally important. | 39 |
| I need the grade, but feedback is more important. | 16 |
| Grade without feedback is almost worthless. | 3 |

*Table 2* and *4* confirm: ***RQ2*** *(Does it increase reflection?)*: 60% responded that they reflected more than when receiving the feedback traditionally; and 67% point out that feedback is for them at least equally important as the grade.

***RQ3*** *(Is the time needed to set up the feedback system limited in comparison to traditional approaches?)* is more difficult to answer firmly, but an estimate is possible. Given 77 project groups and being able to cater 5 sessions per hour would result in almost 16 hours nonstop. Setting up the feedback quiz took about 2 hours, resulting in an 8-factor reduction. Equally important is the fact that the feedback quiz ensures that all students receive their feedback at the same time. Using oral feedback sessions that would need to be fit in the busy schedule of both teachers and students, it would be nearly impossible to deliver the feedback of the voluntary task to all students in time for the consecutive mandatory assignment.

Finally, Cohen and Sasson (2016) observed that results of online quizzes are a good predictor of final exam grades, but in our case, participation sufficed: students doing the voluntary task scored higher than students who did not, rising from 3,63/20 to 5,22/20 in the second year.

## 6. Threats to validity

When comparing the grade for the entire course, the control group consisted of students without doing the task nor the quiz, but we had no control group of students who only missed the quiz. Therefore, we couldn't isolate the effects of the feedback quiz from merely executing the task. Next, participation in the experiment was voluntary and not random. This possibly implies a biased sample, consisting of motivated students who would have scored higher in the course regardless of the feedback quiz, and wo appreciate feedback more than undermotivated students. Nevertheless, the approach scored consistently high and the effect on the grade was more than substantial, hinting at a significant effect of the methodology.

The reduction in time spent with a factor 8, without sacrificing the quality of the feedback, was a very rough estimate and depends on the size of the group and the time needed to setup the feedback quiz, but it is certainly realistic for large student groups.

## 7. Conclusion

We have presented an experiment in providing activating feedback to students through a multiple-choice quiz in the context a voluntary assignment that precedes a larger mandatory task. In total 236 students participated. Approx. 85% said they learned from the feedback and 60% reported reflecting more intensively because of the feedback system, with grades on average 5,22/20 higher than who did not participate.

To be able to differentiate in the effect of performing the assignment and of the feedback methodology, additional qualitative research needs to be done to identify the underlying principles and expand their impact. Care should also be taken when composing the feedback quiz to avoid equity issues related to automated feedback systems which is more present in large classes (Xie et al. (2022)) where the time benefits of this approach are the most present.